\documentclass[prb]{revtex4}
\usepackage{graphicx}
\usepackage{bm}
\usepackage{color}

\begin{document}

\title{Virial relations for ultracold trapped Fermi gases with finite range interactions
through the BCS-BEC crossover}

\author{L. E. C. Rosales-Z\'arate}
\email[]{laura@fisica.unam.mx}
\affiliation{Instituto de F\'{\i}sica, Universidad Nacional Aut\'{o}noma de M\'{e}xico,
Apdo. Postal 20-364, M\'{e}xico D.F. 01000, M\'{e}xico.}
\author{R. J\'auregui}
\email[]{rocio@fisica.unam.mx}
\affiliation{Instituto de F\'{\i}sica, Universidad Nacional Aut\'{o}noma de M\'{e}xico,
Apdo. Postal 20-364, M\'{e}xico D.F. 01000, M\'{e}xico.}
\date{\today}

\begin{abstract}
We study the virial relations  for  ultracold trapped two component Fermi gases
 \cite{Tan,Werner} in the case of short finite range interactions.
Numerical verifications for such relations are reported through the BCS-BEC crossover.
As an intermediate step, it is necessary to evaluate the partial derivatives of the
many body energy with respect to the inverse of the scattering length and with respect
to the interaction range. They are found to have extreme values at the unitary limit.
The virial results are used to check the quality of the variational wave function
involved in the calculations.

\end{abstract}

\maketitle

In the absence of interaction, the virial theorem  relates the energy per
particle of a confined atomic gas with the trapping potential. If that
potential is harmonic, the theorem states that the total energy per particle
is twice the mean trapping potential energy
\begin{equation}
E = 2E_{tr}. \label{virialt}
\end{equation}
For  strongly interacting two component Fermi gases, confined by a harmonic trap
in the unitary limit, this relation was also shown to be valid
experimentally and theoretically \cite{Thomas05, Hu}. The first
derivation of that theorem considered zero-range interactions and made use of the
local density approximation. Further insight on the fundamental
basis of this relation revealed several remarking  features of the
unitary gas such as  its scaling properties \cite{Werner_castin}
or a mapping, using group theory, between the trapped and the free space problem \cite{Mehen}.
Recently, the Hellmann-Feynman theorem was used
to prove Eq.~(\ref{virialt}) at the unitary limit \cite{Son,Thomas08}, and to generalize  the
virial relations for finite scattering lengths
\cite{Tan,Werner}. In fact, general confinement
potentials  and finite range interactions can directly be taken into
account using  a general virial theorem which can be
stated as follows \cite{Werner}:

  Consider a Hamiltonian for a system of N particles with arbitrary statistics:
\begin{equation}\label{gen_ham}
H = H' + U(\vec{r}_1, ..., \vec{r}_N),
\end{equation}
where $H'$ and its domain depend on $p$ parameters with length dimensions
 $\ell_1,...,\ell_p$, on $\hbar$ and the mass of the particles.
$U(\vec{r}_1, ..., \vec{r}_N)$ denotes a regular arbitrary function that
allows  the domains of $H$ and $H'$ to coincide, $\vec{r}_i$ is the
position vector for the $i$-th particle. Then,
\begin{equation}\label{virial_gen}
E = \bigg \langle U + \frac{1}{2} \sum_{i=1}^N \vec{r}_i \cdot \nabla_{\vec{r}_i} U  \bigg \rangle
- \frac{1}{2} \sum_{q=1}^p \ell_q \frac{\partial E}{\partial \ell_q}
\end{equation}
with $E$ the total energy.

For $N$ particles  confined by a harmonic trap, $U=\sum_i^N m(\omega^2_x |x_i|^2 +\omega^2_y |y_i|^2+\omega^2_z |z_i|^2)/2$
and  Eq.~(\ref{virial_gen}) becomes:
\begin{equation}
E = 2E_{tr} - \frac{1}{2} \sum_{s=1}^p \ell_q \frac{\partial
E}{\partial \ell_q}, \label{virial_trapped}
\end{equation}
where $E_{tr}=\langle U\rangle$ is the trapping potential energy.

In the present article, we study $2N$ fermionic atoms in two equally
populated hyperfine states ($N=N_{\uparrow} =N_{\downarrow}=165$)
confined by an isotropic three-dimensional harmonic trap of
frequency $\omega$, and interacting through an attractive finite range
potential $V = - |V_0| e^{-r/r_{\rm v}}$. This potential
is characterized by two parameters, its strength $V_0$ and
its range $r_{\rm v}$. When the kinetic energy of the atoms
is low enough, the scattering length is a proper parameter to
describe the interacting system. For a given number of
$s$-wave bound states and a given $r_{\rm v}$, there is a one-to-one
relationship between the strength of the potential $V_0$
and the scattering length $a$. We consider the case where at
most one bound state is admitted by the potential and find
the ground state of the many body Schr\"odinger equation approximately,
via a variational Monte Carlo calculation, for several scattering lengths $a$ and
short  potential ranges $r_{\rm v}<<r_{ho}\equiv\sqrt{\hbar/m\omega}$. We then
study the behavior of the total, internal and trapping energy as
a function of  both length parameters $a$ and $r_{\rm v}$  to verify Eq.~(\ref{virial_trapped}).
The explicit expression of the Hamiltonian is
\begin{equation}
H=\sum _{i,j=1} ^{N}\frac{p_{\uparrow i}^2 +p_{\downarrow j}^2}{2m}+
  \frac{1}{2}m\omega^2 \left( r_{\uparrow i}^2 + r_{\downarrow j}^2\right)+
\sum_{i,j} V_{\uparrow_i\downarrow_j},
\label{Ham_many_body}
\end{equation}
and the corresponding virial relation becomes
\begin{eqnarray}
E &=& 2E_{tr} - \frac{r_{\rm v}}{2}  \frac{\partial E}{\partial r_{\rm v}}
{\Big|}_{a=constant} - \frac{a}{2}
\frac{\partial E}{\partial a}{\Big|}_{r_{\rm v}=constant}\\
&=& 2E_{tr} - \frac{r_{\rm v}}{2}
\frac{\partial E}{\partial r_{\rm v}}{\Big|}_{a=constant} + \frac{1}{2a}
\frac{\partial E}{\partial (1/a)}{\Big|}_{r_{\rm v}=constant}. \label{virial:b}
\end{eqnarray}

In the BEC side of the crossover the total energy $E$ can become
extremely large compared to the total energy $E$ in the BCS side
 due to the contribution of the binding energy of the formed molecules. This fact
increases the numerical errors in the evaluation of the derivatives in Eq.(\ref{virial:b}).
In order to isolate this two-body effect from many-body effects, we have found
convenient to take into account the  behavior of the free space
 binding energy as follows. The two body problem,
\begin{equation}
\left[\frac{ { p}^2}{2\mu } +V(r)\right]\tilde \varphi({\bf r}) = \varepsilon\tilde\varphi({\bf r}), \quad\quad \mu = m/2,
\label{eq:rel}\end{equation}
is analytically solvable for $s$-states, so that the scattering length is explicitly given by
\begin{eqnarray}
 a &=& r_{\rm v} \eta(\zeta)\nonumber\\
  & =&-2r_{\rm v}\Big[ \frac{\pi}{2} \frac{N_0(\zeta)}
{J_0(\zeta)} - \log( \zeta/2) -C \Big],\label{scatteringlength}
\end{eqnarray}
with $\zeta=(2r_{\rm v}\sqrt{\vert V_0\vert
m}/\hbar)$, $C=0.577215664901...$ is the Euler constant
 and $J_\nu$ and $N_\nu$  represent the Bessel function of the first
and second kind of order $\nu$, respectively.
This problem has the following bound states
\begin{equation}
\tilde\varphi(r) = {\cal N} J_{2r_{\rm v}\sqrt{\vert \varepsilon_s^{(r_{\rm v})} \vert m}/\hbar}(y),
 \label{rarita}
 \end{equation}
where ${\cal N}$ is a normalization factor and $y = \zeta e^{-r/2r_{\rm v}}$.
The boundary condition at the origin implies $J_{x_s}(\zeta) =0$,
so the corresponding  eigenenergies $\varepsilon_s^{(r_{\rm v})}$ fulfill the equation
\begin{equation}\label{energy_cond}
2r_{\rm v}\sqrt{\vert \varepsilon_s^{(r_{\rm v})} \vert m}/\hbar = x_s.
\end{equation}
That is, $x_s$ is determined by $\zeta$ and
\begin{equation}
\varepsilon_s^{(r_{\rm v})} = - \frac{(\hbar x_s)^2}{4mr_{\rm v}^2}.
\end{equation}
We shall work with $z_0<\zeta<z_1$ with $z_0$ and $z_1$
the first two zeros of the Bessel function $J_0$.
Under these conditions,  just one bound state is admitted for each
positive scattering length $a$. Given $a$ and $r_{\rm v}$ and using Eq.~(\ref{scatteringlength}),we can write
\begin{equation}
x_0 =x_0(\zeta) = x_0\Big(\eta^{-1}\Big(\frac{a}{r_{\rm v}}\Big)\Big) \equiv w\Big(\frac{a}{r_{\rm v}}\Big).\label{eq:units}
\end{equation}
 As a consequence, the ground state binding  energy $\varepsilon^{(r_{\rm v})}_0$ of the
 two interacting particle system in otherwise free space satisfies the equation
\begin{eqnarray}\label{free_energy}
r_{\rm v}\frac{\partial \varepsilon^{(r_{\rm v})}_0}{\partial r_{\rm v}}{\Big|}_{a = constant} &=& -2\varepsilon^{(r_{\rm v})}_0 -2\frac{\varepsilon^{(r_{\rm v})}_0 w^\prime(a/r_{\rm v})}{x_0}\frac{a}{r_{\rm v}}\nonumber\\
&=& - 2\varepsilon^{(r_{\rm v})}_0 - a \frac{\partial\varepsilon^{(r_{\rm v})}_0}{\partial a}\Big|_{r_{\rm v}=constant}.
\end{eqnarray}
Thus, if we define
\begin{eqnarray}
\tilde E &=&\frac{E}{2N} \quad \quad\quad\quad\quad{\rm if} \quad a<0, \nonumber\\
         &=&\frac{E}{2N}- \frac{\varepsilon_0^{(r_{\rm v})}}{2}\quad \quad {\rm if} \quad a>0, \label{eq:tildeE}
\end{eqnarray}
and
\begin{equation}
\langle m\omega^2 R^2\rangle =\langle \frac{\sum_i^N m\omega (|r_{\uparrow i}|^2+|r_{\downarrow i}|^2)}{2N}\rangle,
\end{equation}
the virial relation,  Eq.~(\ref{virial:b}), reads
\begin{equation}
\langle m\omega^2 R^2\rangle = \tilde E
+ \frac{r_{\rm v}}{2} \frac{\partial\tilde E}
{\partial r_{\rm v}}{\Big|}_{a= constant} -\frac{1}{2a}
\frac{\partial \tilde E}{\partial (1/a)}{\Big|}_{r_{\rm v} = constant}.
\label{virial:eb}
\end{equation}
This expression is easier to verify numerically than Eq.~(\ref{virial:b}).
Notice that, from a dimensional analysis, equations similar to Eq.~(\ref{eq:units}) can be expected
to be valid for other forms of the potential.

Approximate ground state eigenfunctions for the Hamiltonian Eq.~(\ref{Ham_many_body})
were obtained variationally.
 The trial wave functions used have the  Eagles-Leggett form
\begin{equation}
\Psi_{\lambda_{EL}}= {\cal A}_\uparrow{\cal A}_\downarrow\left[ \phi(1_{\uparrow},1_{\downarrow})
\phi(2_{\uparrow},2_{\downarrow}) ...
\phi(N_{\uparrow},N_{\downarrow})\right]
\label{EL-function}
\end{equation}
through the BCS-BEC crossover regime. In this equation, ${\cal A}$ denotes the antisymmetrizing
operator to be applied to all fermions of each species and
\begin{equation}
\phi({\bf r}_{i\uparrow},{\bf r}_{j\downarrow})\cong
\varphi(r_{i,j})e^{-\lambda_{EL}|{\bf r}_{i\uparrow}+{\bf
r}_{j\downarrow}|^2/4}, \label{eq:becsingle}
\end{equation}
where $\varphi(r_{ij})$ is the s-wave ground state solution of the
trapped interacting two body problem
\begin{equation}
\left[\frac{ { p}^2}{2\mu } + \frac{1}{2}\mu\omega^2 r^2 + V(r)\right] \varphi( r) = \varepsilon\varphi( r).
\label{eq:rel:trapped}
\end{equation}
 The variational parameter $\lambda_{EL}$ modulates the optimal shape
of the atomic cloud. The evaluation of the mean value of the many body Hamiltonian, Eq.(\ref{Ham_many_body}),
for the Eagles-Leggett trial wave function was done using  Monte Carlo techniques
that take advantage of the structure of the function \cite{us08}.

For weak interactions, that is for negative scattering lengths shorter than the mean
separation between interacting atoms, lower variational
energies are obtained using the length scaled ground state solution of the noninteracting problem
(which is a product of Slater determinants) multiplied by a Jastrow correlation function
\begin{eqnarray}
\Psi_{\beta,\lambda_J} &=& F^J_{\lambda_{J}}{\cal A_\uparrow}  \prod_{i=1,N}\phi^{ho}_{{\bf n}_{\uparrow i}} (\beta{\bf r}_{\uparrow i})
              {\cal A_\downarrow}\prod_{i=1,N}\phi^{ho}_{{\bf n}_{\downarrow i}}(\beta{\bf r}_{\downarrow i}),\\
F^J_{\lambda_{J}}&=&\exp [-\lambda_{J}
\sum_{i_\uparrow,j_\downarrow} V(\vert {\bf r}_{i\uparrow}-{\bf r}_
{j\downarrow}\vert )].
\end{eqnarray}
The scaling factors $\beta$ and $\lambda_{J}$ were taken as variational parameters.
For the many body ground state calculation, the inputs of the Slater determinants
are the single-particle eigenstates of the noninteracting trapped system
$\phi^{ho}_{\bf n}({\bf r})$, and the set of  quantum numbers $\{{\bf n}\}$
are chosen to give the lowest energy compatible with Pauli exclusion principle.
The Fermi energy $\epsilon_F$ derived by this procedure can be used to define
an effective Fermi wave number $k_F = \sqrt{2m\epsilon_F}/\hbar$, whose
inverse is a natural unit for measuring the scattering length in the many body problem.

The potential ranges used to perform the calculations were in the interval
$r_{\rm V}/r_{ho} \epsilon [0.002,0.015]$. For each $r_{\rm V}$ and several scattering
lengths through the crossover, upper bounds of the
energy $\tilde E$ were obtained by optimizing the variational parameters $\lambda_{EL}$ or
$\lambda_J$ and $\beta$ according to the trial wave function. The optimization of the
numerical subroutines allowed to explore higher statistics with respect to previous
calculations \cite{us07} and yield the evaluation of the energy with improved accuracy.
In Figures 1 and 2 we illustrate the obtained energies, the vertical size of the plotted
points are  comparable or higher than the numerical errors.
Notice that the resulting  curves for the energy  dependence on the scattering length
for a given range and on the range for a given scattering length show a soft structure that allows a
numerical interpolation or even an analytical local fitting curve.
 These interpolations were used to numerically compute the derivatives necessary to verify the
virial relations. The results are illustrated in Figures 3 and 4. A remarkable feature found
for all the short interaction ranges explored is that both
derivatives, $\partial \tilde E/\partial(1/k_Fa)$ and
$\partial \tilde E/\partial r_{\rm V}$, get an extreme value at unitarity.

In Figure 5, we show a comparison between the trapping energy curve
as predicted by the virial relation, Eq.(\ref{virial:eb}) and specific values of that energy
evaluated directly from the variational functions for a potential range $r_{\rm V}= 0.0025 r_{ho}$.
Notice that there is good overlap between variational
and virial results all over the crossover.
The width of the continuous curve corresponds to the numerical errors in its  derivation.
The main source of error for the trapping energies evaluated directly
from the variational wave functions results from
the non uniqueness of the variational parameters $\lambda_{EL}$ and $\beta$ that yield
similar variational energies. In fact, the agreement between the trapping energy
evaluated directly and using the virial relation can be used as an additional criteria to
select those  parameters.
As a reference, in Fig.~5 the many body energy curve $\tilde E(1/k_Fa)$ is also shown.
The crossing of virial $\langle m\omega R^2\rangle$ and total energy  $\tilde E$
curves does not occur at unitarity as a finite range effect.

Summarizing, we have studied the virial
relations expected for a balanced mixture of two species fermionic  trapped atoms interacting
through a two parameter (intensity and potential range)
attractive potential. We have applied those relations to the study of the ground state
solution of this problem when it is approximately  obtained using simple form
variational trial wave functions. In this way, we are able to quantify the
quality of our wave functions and, even more important, we can
compare the virial relations of short finite range interaction $versus$
contact interactions. Along the calculations, the partial derivatives of the energy $\tilde E$
as a function of the scattering length and of the potential range were numerically evaluated.
It was found that in all cases considered those derivatives get extreme values at
unitarity.  As a consequence, an accurate determination of
the coefficient $\tilde E/E_{IFG}$ at unitarity using finite range potentials require
an accurate extrapolation procedure. In a similar way, an accurate experimental determination of such
coefficient does require an extremely precise realization of the $1/k_Fa\rightarrow 0$ limit.

\begin{figure}
\includegraphics[width=4in]{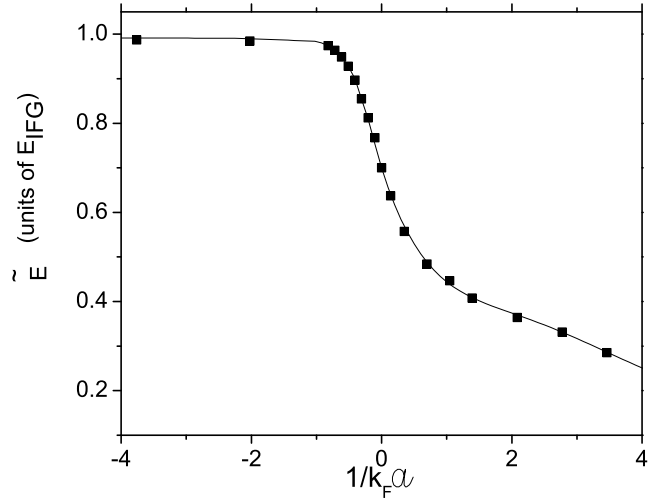}
\caption{ Many body energy $\tilde E$, Eq.(\ref{eq:tildeE}), as a function of the inverse of
the scattering length for a potential range $r_{\rm V}= 0.0025 r_{ho}$
The energy units correspond to $E_{IFG}$ the energy of the non interacting atomic cloud. The
range of the potential is measured in the length scale established by the trapping potential
$r_{ho} = \sqrt{\hbar/m\omega}$; $k_F$ denotes the Fermi wave number.}
\label{fig_1a}
\end{figure}

\begin{figure}
\includegraphics[width=4in]{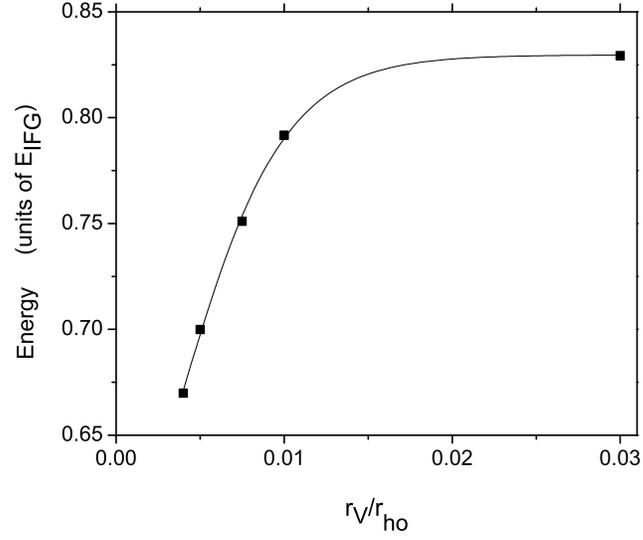}
\caption{  Many body energy  $E$ as a function of the potential range $r_{\rm V}$ at unitarity, $1/k_Fa =0$.
The energy units correspond to $E_{IFG}$, the energy of the non interacting atomic cloud. The
range of the potential is measured in the length scale established by the trapping potential
$r_{ho} = \sqrt{\hbar/m\omega}$.}
\label{fig_1b}
\end{figure}

\begin{figure}
\includegraphics[width=4in]{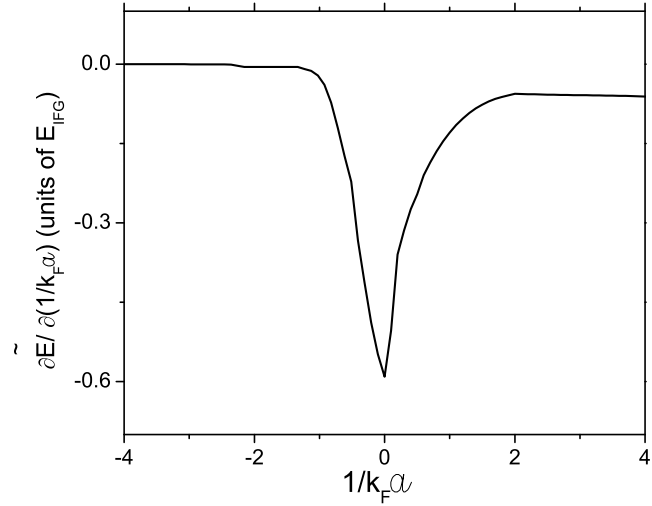}
\caption{ Partial derivative of the many body energy $\tilde E$, Eq.(\ref{eq:tildeE}),
 with respect to the inverse of the scattering length as a function of the inverse of the scattering length
for a potential range $r_{\rm V}= 0.0025 r_{ho}$.
The energy units correspond to $E_{IFG}$, the energy of the non interacting atomic cloud;
$k_F$ denotes the Fermi wave number.}
\label{fig_2}
\end{figure}

\begin{figure}
\includegraphics[width=4in]{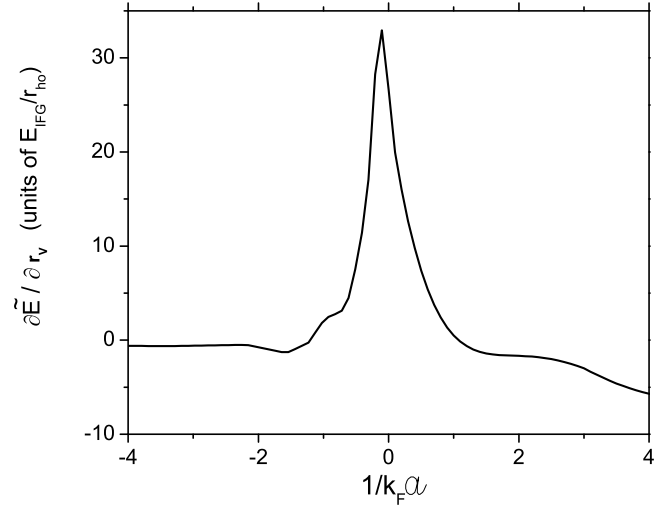}
\caption{ Partial derivative of the many body energy $\tilde E$ with respect to the
potential range $r_{\rm V}$ as a function of the inverse of the scattering length for a potential range $r_{\rm V}= 0.0025 r_{ho}$.
The energy units correspond to $E_{IFG}$, the energy of the non interacting atomic cloud; $k_F$ denotes the Fermi wave number.}
\label{fig_3}
\end{figure}

\begin{figure}
\includegraphics[width=4in]{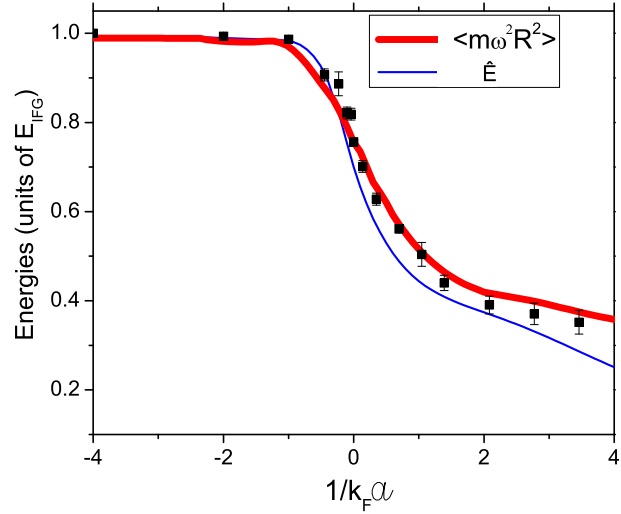}
\caption{(Color online)Softened curves for the many body energy $\tilde E$ and the mean value of twice the trapping energy $<m\omega^2R^2>$
as predicted by the virial relation, Eq.(\ref{virial:eb}). The dots represent the value of the trapping
 energy obtained directly from the variational wave functions. The energy units correspond to $E_{IFG}$,
the energy of the non interacting atomic cloud; $k_F$ denotes the Fermi wave number.} \label{fig_4}
\end{figure}

\end{document}